# The Bright Optical Flash and Afterglow from the Gamma-Ray Burst GRB 130427A

**Authors:** W. T. Vestrand[1]*, J. A. Wren[1], A. Panaitescu[1], P.R. Wozniak[1], H. Davis[1], D. M. Palmer[1], G. Vianello[2], N. Omodei[2], S. Xiong[3], M. S. Briggs[3], M. Elphick[4], W. Paciesas[5], W. Rosing[4]

**Affiliations:**

[1] Los Alamos National Laboratory, P.O. Box 1663, Los Alamos, NM 87545

[2] W.W. Hansen Experimental Physics Laboratory, Kavli Institute for Particle Astrophysics and Cosmology, Department of Physics, and SLAC National Accelerator Laboratory, Stanford University, Stanford, CA 94305

[3] Center for Space Plasma and Aeronomic Research, University of Alabama in Huntsville, 320 Sparkman Dr., Huntsville, AL 35899

[4] Las Cumbres Observatory Global Telescope Network, Inc. 6740 Cortona Drive, Suite 102, Santa Barbara, CA 93117

[5] Universities Space Research Association, 320 Sparkman Dr., Huntsville, AL 35899

*Correspondence to: vestrand@lanl.gov

**Abstract:** The optical light that is generated simultaneously with the x-rays and gamma-rays during a gamma-ray burst (GRB) provides clues about the nature of the explosions that occur as massive stars collapse to form black holes. We report on the bright optical flash and fading afterglow from powerful burst GRB 130427A and present a comparison with the properties of the gamma-ray emission that show correlation of the optical and >100 MeV photon flux light curves during the first 7,000 seconds. We attribute this correlation to co-generation in an external shock. The simultaneous, multi-color, optical observations are best explained at early times by reverse shock emission generated in the relativistic burst ejecta as it collides with surrounding material and at late times by a forward shock traversing the circumburst environment. The link between optical afterglow and >100 MeV emission suggests that nearby early peaked afterglows will be the best candidates for studying particle acceleration at GeV/TeV energies.

**Main Text:** Long-duration gamma-ray bursts are associated with the collapse of massive stars to form black holes (1) or rapidly-spinning, highly-magnetized, neutron stars (2). This collapse is believed to eject collimated relativistic jets that, through internal dissipation processes and collisions with the surroundings, generate luminous outbursts of electromagnetic radiation that have been detected at radio frequencies to very high (GeV) gamma-ray energies. Most of the outburst energy is emitted in the gamma-rays. But, starting with the first observations that established that GRBs occur at cosmological distances (3), correlative optical observations, in particular, have proven themselves as important tools for unraveling the nature of GRB explosions. Here we present observations of the optical flash and early afterglow for a nearby burst that is bright enough in very high energy gamma-rays to allow a detailed comparison of the >100 MeV gamma-ray and optical light curves. These optical observations cover the critical early phases of the explosion from the time interval before the event onset, through the bright optical and prompt gamma-ray emitting period, and well into the early afterglow phase.

Starting at 27 April 2013 at 07:47:06.42 UTC (hereafter $T_o$), the Gamma Ray Burst Monitor (GBM) on the Fermi Satellite, the Burst Alert Telescope (BAT) on the Swift Satellite, and an armada of other space-based gamma-ray detectors detected the onset of a powerful gamma-ray burst (GRB) (4,5). This GRB, called GRB 130427A, had the largest gamma-ray fluence (~$2.7 \times 10^{-3}$ erg/cm$^2$ in the 20 keV-1200 keV band) measured in more than 18 years of operation by Konus-Wind (6) and set a record for duration of the >100 MeV gamma-ray emitting interval (5). Spectroscopy of the optical counterpart (7), coarsely localized by the Swift BAT and later refined by follow-up with optical telescopes, places the GRB at a redshift z=0.34--- a distance about five times closer than the typical GRB localized by Swift. However, even accounting for its proximity, the intense gamma-ray fluxes observed imply an apparent isotropic energy release of nearly $10^{54}$ ergs and rank it among the more powerful GRBs ever detected (3). Subsequent optical monitoring discovered the emergence of a broad-line supernova at the GRB location (8).

This powerful GRB also generated an extremely bright flash of optical emission and a long-lived, bright, optical afterglow. Three independent RAPTOR (RAPid Telescopes for Optical Response) full sky monitoring telescopes (9), at locations in New Mexico and Hawaii, detected the emergence of a bright flash, temporally coincident with the onset of gamma-ray emission, at the location of GRB 130427A. The optical flash rapidly peaked at a magnitude 7.03±0.03 (unfiltered observations calibrated to Sloan r' band) in an exposure that covered the time interval $T_o$+9.31 to $T_o$+19.31 seconds. After the peak, the flash faded with a power-law flux decay with index α=-1.67±0.07 ($\chi^2$=0.68/5 dof) and was detected for about 80 seconds until it faded below the ~10th magnitude sensitivity limit of the RAPTOR full sky monitors.

The taxonomy for optical emission detected during the prompt gamma-ray emitting interval identifies two broad classes: prompt optical emission correlated with prompt gamma-ray emission (10-12) and early optical afterglow emission uncorrelated with the prompt gamma-ray emission (11,13,14). In context of the standard fireball model (15, 16), the prompt optical emission is attributed to internal shocks in an ultra-relativistic jet outflow generated by the central engine and the afterglow emission to external shocks generated by interaction with the surrounding medium. The prompt optical emission therefore reflects the impulsive energy injection into the jet and the early afterglow emission measures the response of the jet/environment system to the energy injection. Bright optical flashes from reverse shocks were predicted on theoretical grounds (16,17) before observational evidence was seen in GRB 990123 (13). The optical flash light-curve for GRB 130427A shows a single peak delayed with respect to the keV-MeV prompt gamma-ray peak (Fig. 1) and a steep power-law flux that is consistent with the predictions of models for optical flashes from reverse shocks (17). Based on the above taxonomy, the brightness of the flash, and the rapid power law flux decay, it makes sense to associate the optical flash with reverse shock emission.

To explore the evolution of the broad band GRB spectrum during the optical flash, we constructed spectral energy distributions (SED) using simultaneous measurements taken with the

Fermi GBM and the Fermi Large Area Telescope (LAT). Each snapshot of the time evolving SED was formed by integrating the GRB flux over the same time interval as the optical exposure. We found that the broad-band SEDs (Fig. 2) varied rapidly during the first 40 seconds and the optical measurements fell far from the values expected from extrapolation of the keV-MeV SED. However as the intensity of the outburst declined during the next 40 second interval, the SED shape stabilized and the optical measurements started to converge on the values predicted by a straightforward linear extrapolation of the keV-MeV SED. By the end of the optical flash, the optical to 10 MeV spectrum is consistent with a single power law with index $\beta$=-0.64.

In response to the Swift BAT localization alert at 127.8 seconds after the GBM trigger, our RAPTOR response telescopes began unfiltered and simultaneous multicolor (g', r', i', z') optical observations at $T_o$+132.9 seconds that continued until $T_o$+7,585.9 seconds. This photometry begins near the peak of a prominent flare in keV-MeV x-ray/gamma-rays that lasts until ~$T_o$+400 seconds. The optical light curves show a smooth monotonic decline but no indication of the steep decline nor the break to a slower power-law decay at ~400 seconds measured at x-ray energies (4). Instead, the structure of the optical light-curve shows a steepening at about $T_o$+270 seconds. This steepening is essentially achromatic and the color of the optical emission is consistent with a $\nu^{-0.70\pm0.05}$ spectrum and constant until it starts to become bluer ($\nu^{-0.59\pm0.05}$) at ~3,000 seconds after the GBM trigger (see bottom panel of figure 3).

In marked contrast with the keV-MeV emission, the optical light curves after $T_o$+100 seconds show a striking similarity with the >100 MeV photon flux light curve measured by the Fermi LAT (5). The LAT light curve has a break at about 300 seconds, just like the optical afterglow. Straightforward scaling of the RAPTOR optical light curve by a factor of ~$10^{-6}$ provides a reasonable description of the LAT observations out to ~$T_o$+7000 seconds. This close correspondence argues for a common origin of both components in external shocks.

The optical light curve until ~$T_o$+3,000 seconds is best modeled by synchrotron emission from a reverse shock in a wind density profile. Most optical afterglows have been modeled with forward shocks in a homogeneous medium. But the peak brightness (~6 Jy) and steep decay of the optical flash suggest origin in a reverse shock. The relative faintness of the radio afterglow peak (~1 mJy) also argues for generation by a reverse shock in a wind-like medium (18). To explain the optical flash by reverse shock emission in a wind ($R^{-2}$) requires either a long-lived electron energy injection up to ~40 seconds or a shorter (~20 seconds) followed by adiabatic cooling. Figure 4 shows the best fit to the optical flash with the short interval of injection (on at $T_o$+4 seconds and off at $T_o$+20 seconds) and, for self-consistency, the same dynamical parameters that we infer from fits to the later afterglow forward shock emission discussed below. This model employs an electron distribution with power-law energy index p=1.88 and corresponds to an injected energy of $8.0 \times 10^{53}$ ergs. The slower optical fading after the flash interval requires a second episode of energy injection to sustain the optical afterglow or a continuous outflow with a variable Lorentz factor (19). This sustained reverse shock model reproduces the closely tracking variability observed by the RAPTOR telescopes and the Fermi LAT and suggests that the optical and most of the >100 MeV emission is generated by synchrotron emitting electrons that are accelerated by the reverse shocks.

This reverse shock model cannot, by itself, explain the properties of the prompt keV-MeV emission nor some of the properties of the late time afterglows. The evolution to a bluer color after ~3,000 seconds observed by RAPTOR and the slowing of the optical brightness decay suggests the emergence of a forward shock component. This transition to forward shock dominance at late times would also naturally explain the late time x-ray light-curve and the sustained >100 MeV emission after 10,000 seconds. Emergence of a bluer optical component at late time is similar to the afterglow evolution of GRB 080319B--- another burst with a bright optical flash. For GRB 080319B, the color change was also interpreted as marking the transition from reverse shock emission dominance to forward shock dominance (12, 20, 21).

During most of the interval before $T_o+400$ seconds, the keV-MeV x-ray/gamma-ray emission is consistent with the standard assumption that the prompt emission is generated by internal shocks in the relativistic jet ejecta. Our predicted >100 MeV flux from the reverse shock that generates the optical flash is slightly less (~ factor of 2) than the peak measured by the Fermi LAT (Fig 4). But the keV-MeV flux is significantly underpredicted by at least a factor of 10. So the reverse shock in a wind model requires prompt emission to explain the keV-MeV emission and might require additional >100 MeV emission to explain the LAT light-curve peak. In this picture, the keV-MeV light-curve is a proxy that traces the injection of internal jet energy by the central engine. The keV-MeV emission therefore indicates two periods of significant energy injection into the jet: the initial 20 seconds and a period from ~120 to 300 seconds. The interesting potential exception to prompt emission dominance in the keV-MeV range is the period just before onset of the flare at $T_o+120$ seconds. During the interval $T_o+79$ seconds to $T_o+89$ seconds, the optical afterglow flux measured by RAPTOR falls right on the extrapolation of the power-law (index=~-0.6) measured in the 10 keV-20 MeV energy band by the GBM. This gamma-ray spectral slope is also similar to spectral slope that we measure for optical afterglow emission at later times. The "notch" in the keV-MeV light curve before the flare at $T_o+120$ may be providing a rare glimpse, similar to that seen in GRB 980923 (22), of afterglow emission at MeV energies between prompt emission intervals.

The exceptional optical properties observed for the optical flash and afterglow from GRB 130427A are mostly a result of burst proximity. The flash peak luminosity for GRB 130427A is among the most powerful events, but its value is consistent with the anti-correlation between peak time and peak luminosity found for optical afterglows (23). If optical afterglows and >100 MeV gamma-ray afterglows have a common origin, then the peaked optical afterglows that peak early should be the best candidates for detection at GeV/TeV gamma-ray energies.

**References and Notes:**

1. S.E. Woosley and J. S. Bloom, Ann. Rev. Astron. Astrophys., **44,** 507 (2006).
2. B. D. Metzger et al., Mon. Not. Royal Astron. Soc., **413**,2031 (2011).
3. M. R. Metzger et al., Nature, **387**, 879 (1997).
4. A. Maselli et al., this issue of Science.
5. The Fermi Team, this issue of Science.
6. S. Golenetskii et al. GCN Circular#14486 (2013).
7. A. J. Levan et al., GCN circular #14456 (2013).
8. D. Xu et al., Astrophysical J. in press (2013).
9. J. Wren et al. SPIE, **7737**, 773723 (2010).
10. W. T. Vestrand, et al. Nature, **435,** 178 (2005).
11. W. T. Vestrand et al. Nature, **442**, 172 (2006).

12. J. L. Racusin et al. Nature, **455,** 183 (2008).
13. C. Akerlof et al. Nature, **398**, 400 (1999).
14. E. Rykoff et al. Astrophysical J., **702**, 489 (2009).
15. P. Meszaros and M. Rees, Astrophysical J., **405**, 278 (1993).
16. P. Meszaros and M. Rees, Astrophysical J., **476**, 232 (1997).
17. R. Sari and T. Piran, Astrophysical J., **520**, 641 (1999).
18. T. Laskar et al., Astrophysical J., **776**, 119 (2013).
19. Z. L. Uhm et al. Astrophysical J., **761**, 147 (2012).
20. P. Wozniak et al. Astrophysical J. , **691**, 495 (2009).
21. J.S. Bloom et al. Astrophysical J., **691**, 723 (2009).
22. T.W. Giblin et al. Astrophysical J., **524**, L47 (1999).
23. A. Panaitescu and W.T. Vestrand, MNRAS, **387**, 497 (2008).

**Acknowledgments:** This gamma-ray burst research was supported by NASA and the Laboratory Directed Research and Development program at Los Alamos National Laboratory. The optical measurements reported in this paper are available on-line in the Supplementary Materials.

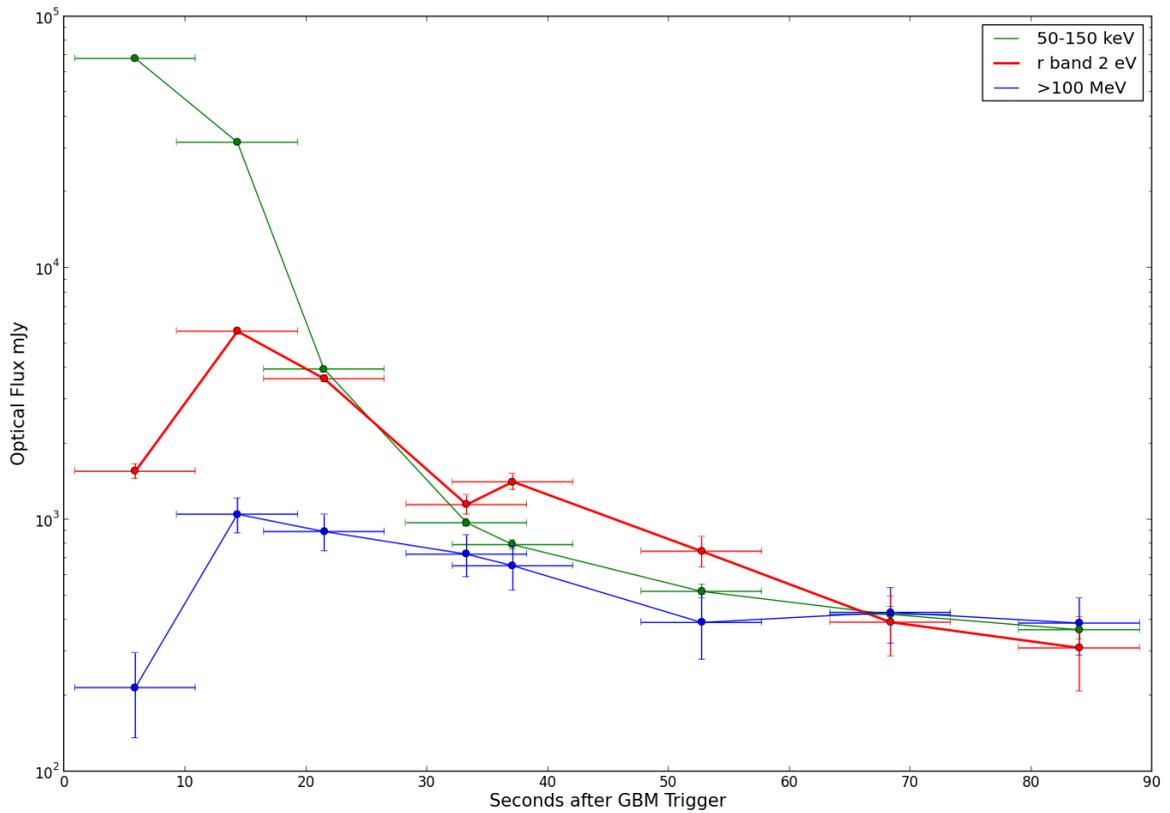

**Fig. 1.** A comparison of the relative flux variations measured for GRB 130427A by the Fermi LAT, RAPTOR, and the Swift BAT during the first 90 seconds after the gamma-ray burst trigger. The >100 MeV emission and the 50-150 keV emission have been integrated over the same time intervals as the optical exposures and multiplied by a scaling factor to allow comparison with the optical light curve. Both the optical and >100 MeV light curves rise to a peak in the second interval, 50-150 keV emission peaks in the first interval.

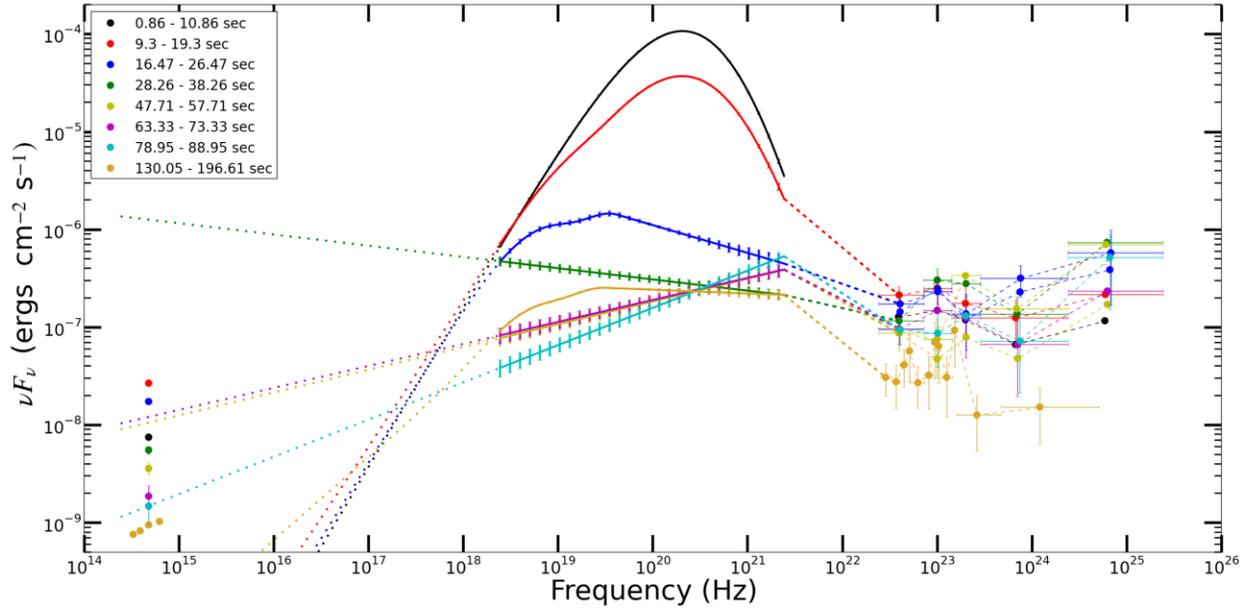

**Fig 2.** The spectral energy distributions measured for GRB 130427A during the early phases of the burst development. The points and solid lines represent actual measurements. The dotted straight lines indicate the extrapolation the keV-MeV measurements for comparison to the optical measurements. The dashed lines indicate the connection between energy bands and are not actual measurements. The measurements were obtained by RAPTOR, the Fermi GBM, and the Fermi LAT. The errors bars on the GBM measurements have been increased by 25% to allow for systematics, such as pulse pile-up (PPU), background subtraction during and following the repointing of Fermi, and the rapid spectral evolution during the exposure intervals.

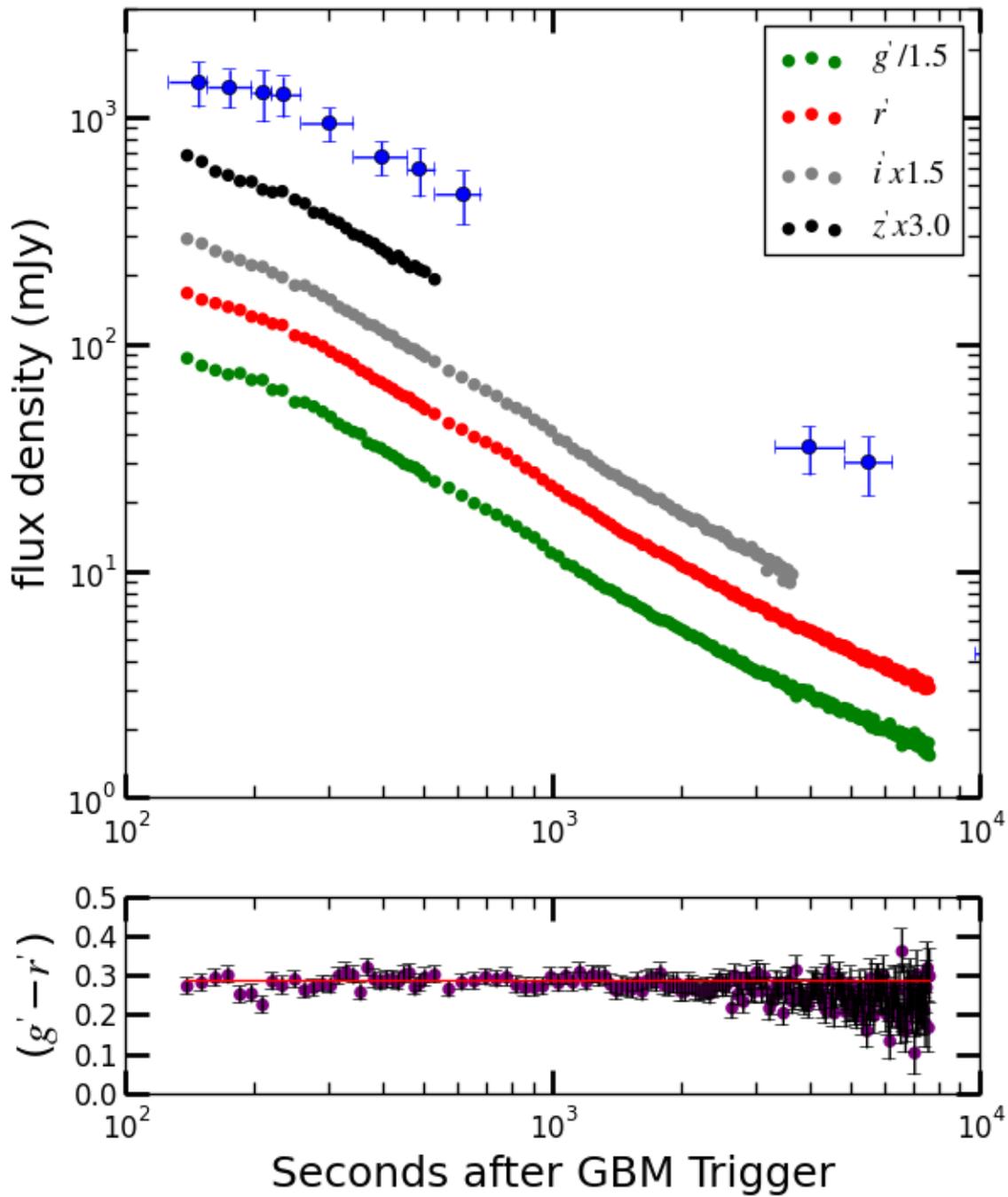

**Fig. 3.** Simultaneous multi-color measurements of the early optical afterglow from GRB 130427A. The top panel shows the light-curves obtained using standard Sloan g', r', I', z' filters. The light-curves show more structure than is captured by power law fits. But the r' band light-curve can be characterized as a power-law decay with index ~0.7 before 270 seconds, ~1.1

between 270 and 3,000 seconds, and 0.88 between 3,000 and 7,500 seconds. Also plotted for comparison, in blue, is the photon flux light curve for >100 MeV emission measured by the Fermi LAT (4). The bottom panel shows the g'-r' color evolution of GRB 130427A. The red line indicates the mean g'-r' value (0.286±0.018) measured before $T_0$+1000 seconds. After about T0+3,000 seconds a bluer component emerges as the flux decay slows.

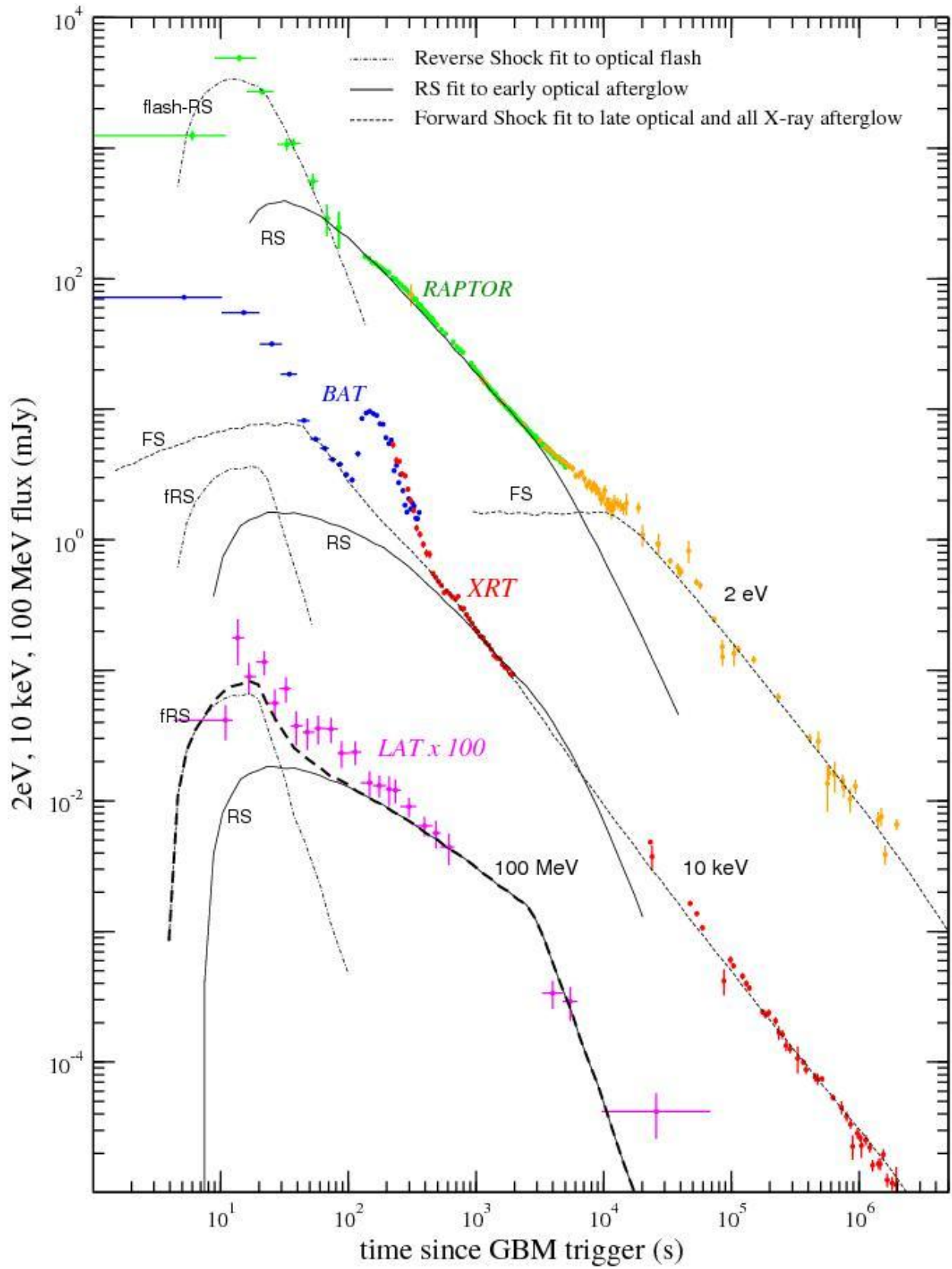

Fig4. Best-fit with a reverse-forward shock model to the optical light-curve of GRB afterglow 130427A. Three episodes of ejecta and energy injection are needed to explain the optical flash, the early optical emission (up to few ks), and the late optical emission (after a few ks). Here each "injection episode" represents a change in the dynamical and micro-physical parameters of the reverse shock in an otherwise continuous ejecta injection into the reverse-shock. For the first two episodes, the optical emission arises from the reverse shock, and the kinetic energy of the incoming ejecta ($6.10^{52}$ erg/sr and $4.10^{53}$ erg/sr, respectively) is less than that of the leading shock ($10^{54}$ erg/sr). During the last injection episode, the optical afterglow emission arises from the forward-shock, with the incoming ejecta carrying $3.10^{54}$ erg/sr, which is more than that already existing in the forward-shock (thus its deceleration is mitigated by the energy injection). Other parameters are: 1) first injection episode, flash reverse-shock (fRS)-- onset at 4 s, end at 15 s, incoming ejecta Lorentz factor 730, magnetic field parameter 0.008, electron energy parameter 0.006, index of electron power-law distribution with energy 1.9 2) second injection episode, reverse-shock (RS, parameters in same order as above) --15 s, 3 ks, 1800, 0.0010, 0.012, 2.0. 3) third injection episode, forward-shock (FS) -- 3 ks, > 2 Ms, > 100, $3.10^{-4}$, 0.14, 2.3, energy injection law $E \sim t^{0.3}$. The micro-physical parameters for the forward shock are kept constant throughout the entire course of the event. We also require that the ambient medium have a wind-like density stratification ($n \sim r^{-2}$) corresponding to a GRB progenitor with a mass-loss rate--to--wind speed ratio of $4.10^{-11}$ (M_sun/yr)/(km/s). Dot-dash lines show the model fits for the flash phase emission, the solid lines show the reverse shock contributions during the early afterglow phase, and the dashed lines shows the contributions of the forward shock emission to the late optical afterglow and all phases of the x-ray afterglow. The reverse-shock emission during the third injection episode is not shown and would be dimmer than that of the forward-shock, if the reverse-shock has the same micro-physical parameters as the forward-shock.

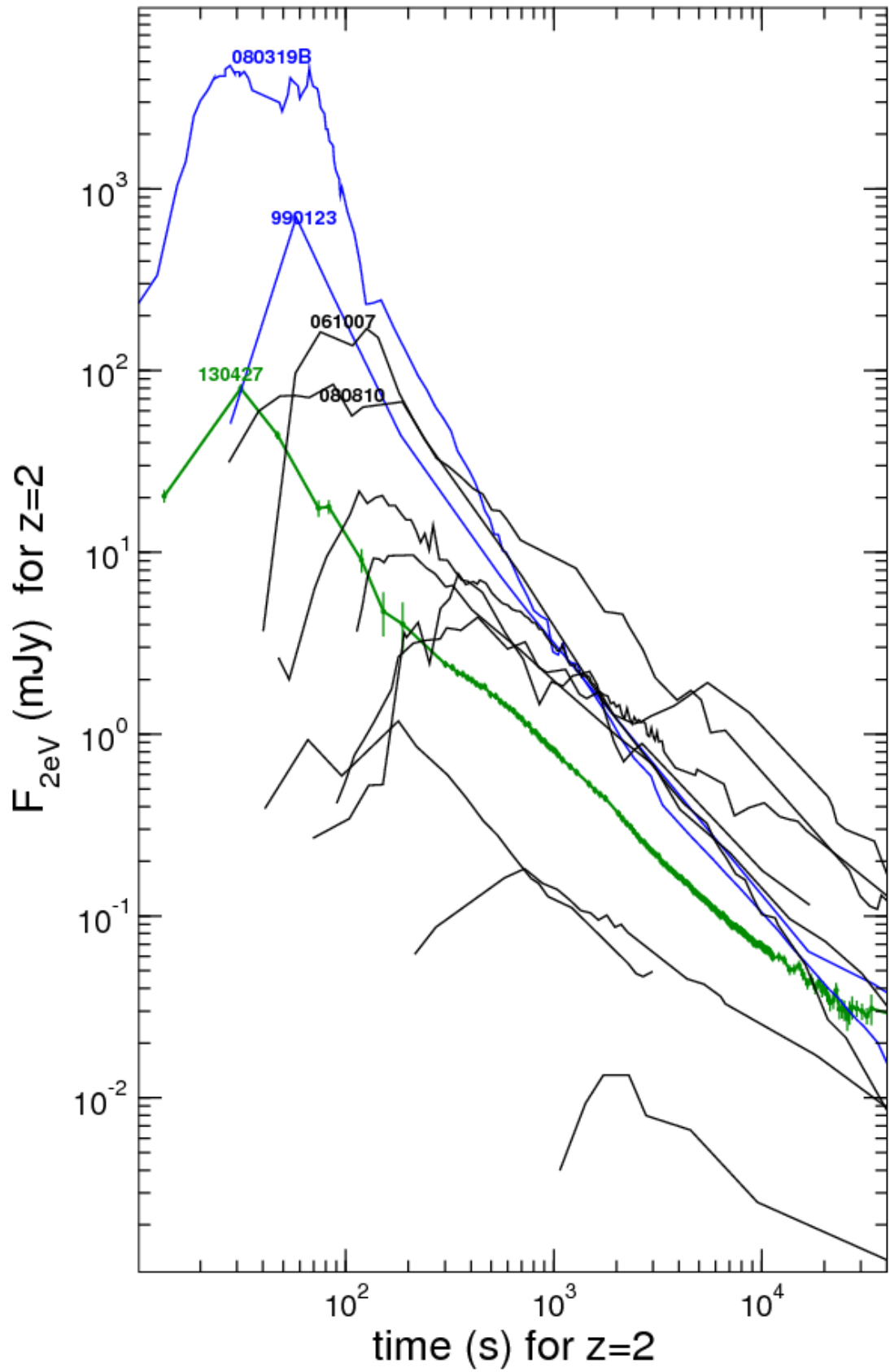

**Fig. 5.** A comparison of the light-curve from GRB 130427A (in green) with those measured for other prominent GRBs with peaked optical afterglows. All of the light-curves have been transformed to show how they would appear if they had occurred at redshift z=2.

**Supplementary materials:**

**Modeling:**

We modeled the GRB 130427A multi-wavelength afterglow emission using the standard external-shock framework (1), where a relativistic forward-shock is driven into the ambient medium by GRB ejecta and that interaction also drives a reverse shock through the ejecta. The 1-dimensional dynamics of each shock is calculated from conservation of mass, energy, and momentum, using the shock jump-conditions (2). The post-shock typical electron energy and magnetic field are calculated assuming that they acquire a certain fraction (micro-physical parameters) of the dissipated energy. Each shock produces synchrotron and inverse-Compton emissions, which are then integrated over the propagation of the shock. The self-absorption and cooling frequencies of the synchrotron spectrum are also calculated from the electron distribution and magnetic field. The best-fit to the afterglow data is then obtained by minimization of chi-square between the sum of model emission components and the observations. For the best fitting models, the effects of synchrotron self-absorption were found to not be important at the observed frequencies during the early epochs. Further, the self-Compton flux from the shocks were also found to be are smaller at all times and all frequencies than the corresponding synchrotron fluxes shown in figure 4. However, we did find that the inverse Compton process does play an important role because dominates the electron radiative cooling and determines the synchrotron efficiency above the 100 MeV cooling frequency. The cross-components of synchrotron photons produced by one shock and Compton up-scattered in the other have not been not calculated, but we expect them to be typically smaller than the joint synchrotron/self-Compton flux from the shock.

Our models that successfully fit the multi-wavelength emission require a reverse-shock that is sustained by energy injection throughout the prompt and early afterglow phases. The nature of this injection must vary in three episodes that signal a change in the dynamical and micro-physical parameters of the reverse shock. The best fit parameters for the initial time interval are flash reverse-shock (fRS)-- onset at 4 s, end at 15 s, incoming ejecta Lorentz factor 730, magnetic field parameter 0.008, electron energy parameter 0.006, index of electron power-law distribution with energy 1.9. To account for the transition from the rapid optical flash decay to the more gradual optical afterglow decay observed after 100 seconds requires a change on the reverse-shock dynamics and micro-parameters at ~15 seconds. After each change, we calculate the adiabatic and radiative cooling of the ejecta electrons of the previous injection episode. When those electrons cool to an energy below that for radiating synchrotron at 100 MeV, the 100 MeV flux from the 1/Gamma fluid moving toward the observer would display an exponential cutoff in time. However, the fluid moving at larger angles yields a brighter flux after the exponential cutoff, and shows a steep power-law flux decay. The steep power-law decay of the "large angle emission" is seen at 100 MeV after 20 s and after 3 ks in Figure 4. We find the best fit parameters for this new interval are reverse-shock energy injection starting at ~15 s and continuing to ~ 3 ksec with a Lorentz factor of 1800, magnetic field parameter of 0.0010, electron energy parameter of 0.012, and electron power-law index of 2.0.

**Observations:**
The three RAPTOR All-Sky Monitors were the first to detect the optical emission from GRB130427A. The burst location was visible at both the Maui and Los Alamos sites at an elevation of 80° and 53° respectively. Each All-Sky Monitor consists of five unfiltered 24 mm f1.5 Canon EF lenses combining to cover about 90% of the sky above 12° elevation. Individually, each All-Sky Monitor obtains a 10 s exposure every 20 s with staggered observing schedules to provide better temporal coverage as a system. We corrected the All-Sky Monitor images using bias and dark frames obtained later that evening as well as flat fields which were created using a median of twilight frames obtained during cloudless evenings during the first half of 2013. The corrected images were then reduced to object lists using the SExtractor package (3) and astrometrically calibrated using ~8600 common field stars from the Tycho-2 (4) catalog with SDSS r' band synthetic magnitude estimates (5). Due to sky gradients caused by a nearly full moon about 75° away, these object lists were then photometrically re-calibrated using ~55 common Tycho-2 stars within 3° of the burst location.

The narrow field instruments located at the Fenton Hill Observatory in Northern New Mexico, RAPTOR-S and RAPTOR-T, began observing the burst location shortly after receiving the Swift trigger. The RAPTOR-S system consists of a single unfiltered 0.4 m telescope and the RAPTOR-T system consists of four co-aligned 0.4 m telescopes obtaining simultaneous images is four photometric bands (SDSS g', r', i', and z'). Our narrow field observations began at $T-T_o$ =132.96 s and the GRB counterpart was clearly visible in all five telescopes. The response sequence of each telescope consists of nine 5 s images followed by twenty 10 s images and 30 s images thereafter. The response images were calibrated using bias and dark frames obtained later that evening as well as flat frames created using a combination of recent twilight images and sky patrol images. The images were reduced to object lists using the SExtractor package and then calibrated using ~60 common field stars from the appropriate band in the SDSS DR9 catalog (6). The unfiltered RAPTOR-S data were calibrated to the SDSS r band values.

Our spectral energy distributions (SEDs) in the optical energy range were constructed using the g', r', i', and z' fluxes averaged over four time intervals with end points at 138, 270, 550, 3000, and 7571 seconds after the burst. We applied flux corrections based on extinction curves in Pei (7) to account for $E_{B-V} = 0.0202$ of Galactic reddening (8) and an $A_V = 0.18$ dust column in the host (9). The best fit power law spectrum has a mean slope of β = -0.70±0.05 until about 3000 seconds after the burst, when it becomes slightly bluer (β = -0.59±0.05).

| $T_{mid}$ (s) | $t_{exp}$ (s) | $C_r$ (mag) | g' (mag) | r' (mag) | i' (mag) | z' (mag) | Telescope (name) |
|---|---|---|---|---|---|---|---|
| colspan=8 | RQD2—All Sky Monitors |||||||
| -9.75 | 10.00 | > 10.1 | ..... | ..... | ..... | ..... | D03 |
| -4.33 | 10.00 | > 9.9 | ..... | ..... | ..... | ..... | D02 |
| 1.81 | 10.00 | > 9.3 | ..... | ..... | ..... | ..... | D01 |
| 5.86 | 10.00 | 8.421+.075-.070 | ..... | ..... | ..... | ..... | D03 |
| 14.31 | 10.00 | 7.030+.030-.029 | ..... | ..... | ..... | ..... | D02 |
| 21.48 | 10.00 | 7.501+.032-.031 | ..... | ..... | ..... | ..... | D03 |
| 22.21 | 10.00 | 7.625+.061-.058 | ..... | ..... | ..... | ..... | D01 |
| 33.27 | 10.00 | 8.752+.098-.090 | ..... | ..... | ..... | ..... | D02 |
| 37.09 | 10.00 | 8.525+.081-.075 | ..... | ..... | ..... | ..... | D03 |
| 52.71 | 10.00 | 9.215+.151-.133 | ..... | ..... | ..... | ..... | D03 |
| 68.33 | 10.00 | 9.919+.291-.229 | ..... | ..... | ..... | ..... | D03 |
| 83.95 | 10.00 | 10.177+.356-.268 | ..... | ..... | ..... | ..... | D03 |
| 99.56 | 10.00 | > 10.0 | ..... | ..... | ..... | ..... | D03 |
| 115.63 | 10.00 | > 10.1 | ..... | ..... | ..... | ..... | D03 |
| 131.25 | 10.00 | > 10.1 | ..... | ..... | ..... | ..... | D03 |
| colspan=8 | RAPTOR-S |||||||
| 135.46 | 5.00 | 10.713+-0.011 | ..... | ..... | ..... | ..... | |
| 144.56 | 5.00 | 10.775+-0.011 | ..... | ..... | ..... | ..... | |
| 153.66 | 5.00 | 10.854+-0.011 | ..... | ..... | ..... | ..... | |
| 162.76 | 5.00 | 10.874+-0.011 | ..... | ..... | ..... | ..... | |
| 171.86 | 5.00 | 10.923+-0.011 | ..... | ..... | ..... | ..... | |
| 180.86 | 5.00 | 10.949+-0.011 | ..... | ..... | ..... | ..... | |
| 189.96 | 5.00 | 10.999+-0.011 | ..... | ..... | ..... | ..... | |
| 199.06 | 5.00 | 11.021+-0.012 | ..... | ..... | ..... | ..... | |
| 208.16 | 5.00 | 11.048+-0.012 | ..... | ..... | ..... | ..... | |
| 221.96 | 10.00 | 11.135+-0.011 | ..... | ..... | ..... | ..... | |
| 234.76 | 10.00 | 11.177+-0.011 | ..... | ..... | ..... | ..... | |
| 247.96 | 10.00 | 11.245+-0.011 | ..... | ..... | ..... | ..... | |
| 261.17 | 10.00 | 11.290+-0.011 | ..... | ..... | ..... | ..... | |
| 273.96 | 10.00 | 11.341+-0.011 | ..... | ..... | ..... | ..... | |
| 287.16 | 10.00 | 11.382+-0.011 | ..... | ..... | ..... | ..... | |
| 299.97 | 10.00 | 11.445+-0.011 | ..... | ..... | ..... | ..... | |
| 313.16 | 10.00 | 11.486+-0.011 | ..... | ..... | ..... | ..... | |

| | | | | | | |
|---|---|---|---|---|---|---|
| 325.97 | 10.00 | 11.558+-0.011 | ..... | ..... | ..... | ..... |
| 338.76 | 10.00 | 11.596+-0.011 | ..... | ..... | ..... | ..... |
| 351.96 | 10.00 | 11.631+-0.012 | ..... | ..... | ..... | ..... |
| 364.76 | 10.00 | 11.664+-0.012 | ..... | ..... | ..... | ..... |
| 377.56 | 10.00 | 11.721+-0.012 | ..... | ..... | ..... | ..... |
| 390.36 | 10.00 | 11.773+-0.012 | ..... | ..... | ..... | ..... |
| 403.16 | 10.00 | 11.814+-0.012 | ..... | ..... | ..... | ..... |
| 416.36 | 10.00 | 11.853+-0.012 | ..... | ..... | ..... | ..... |
| 429.16 | 10.00 | 11.877+-0.012 | ..... | ..... | ..... | ..... |
| 441.96 | 10.00 | 11.917+-0.012 | ..... | ..... | ..... | ..... |
| 454.76 | 10.00 | 11.949+-0.012 | ..... | ..... | ..... | ..... |
| 467.56 | 10.00 | 12.010+-0.012 | ..... | ..... | ..... | ..... |
| 492.56 | 30.00 | 12.049+-0.011 | ..... | ..... | ..... | ..... |
| 534.86 | 30.00 | 12.147+-0.011 | ..... | ..... | ..... | ..... |
| 577.16 | 30.00 | 12.222+-0.011 | ..... | ..... | ..... | ..... |
| 619.56 | 30.00 | 12.294+-0.011 | ..... | ..... | ..... | ..... |
| 661.96 | 30.00 | 12.380+-0.011 | ..... | ..... | ..... | ..... |
| 704.26 | 30.00 | 12.469+-0.011 | ..... | ..... | ..... | ..... |
| 746.66 | 30.00 | 12.526+-0.012 | ..... | ..... | ..... | ..... |
| 788.96 | 30.00 | 12.579+-0.012 | ..... | ..... | ..... | ..... |
| 831.76 | 30.00 | 12.663+-0.012 | ..... | ..... | ..... | ..... |
| 874.16 | 30.00 | 12.725+-0.012 | ..... | ..... | ..... | ..... |
| 916.46 | 30.00 | 12.789+-0.012 | ..... | ..... | ..... | ..... |
| 958.46 | 30.00 | 12.858+-0.012 | ..... | ..... | ..... | ..... |
| 1001.16 | 30.00 | 12.916+-0.012 | ..... | ..... | ..... | ..... |
| 1043.16 | 30.00 | 12.993+-0.012 | ..... | ..... | ..... | ..... |
| 1085.56 | 30.00 | 13.034+-0.012 | ..... | ..... | ..... | ..... |
| 1128.37 | 30.00 | 13.097+-0.012 | ..... | ..... | ..... | ..... |
| 1170.86 | 30.00 | 13.154+-0.012 | ..... | ..... | ..... | ..... |
| 1213.16 | 30.00 | 13.178+-0.013 | ..... | ..... | ..... | ..... |
| 1255.96 | 30.00 | 13.237+-0.013 | ..... | ..... | ..... | ..... |
| 1297.96 | 30.00 | 13.279+-0.013 | ..... | ..... | ..... | ..... |
| 1339.96 | 30.00 | 13.318+-0.013 | ..... | ..... | ..... | ..... |
| 1382.76 | 30.00 | 13.357+-0.013 | ..... | ..... | ..... | ..... |
| 1425.16 | 30.00 | 13.379+-0.013 | ..... | ..... | ..... | ..... |
| 1467.46 | 30.00 | 13.446+-0.013 | ..... | ..... | ..... | ..... |
| 1509.76 | 30.00 | 13.471+-0.013 | ..... | ..... | ..... | ..... |
| 1552.36 | 30.00 | 13.493+-0.013 | ..... | ..... | ..... | ..... |
| 1594.76 | 30.00 | 13.535+-0.013 | ..... | ..... | ..... | ..... |
| 1637.26 | 30.00 | 13.560+-0.013 | ..... | ..... | ..... | ..... |
| 1679.66 | 30.00 | 13.593+-0.013 | ..... | ..... | ..... | ..... |
| 1721.96 | 30.00 | 13.626+-0.014 | ..... | ..... | ..... | ..... |
| 1764.66 | 30.00 | 13.652+-0.014 | ..... | ..... | ..... | ..... |
| 1806.96 | 30.00 | 13.666+-0.014 | ..... | ..... | ..... | ..... |
| 1849.37 | 30.00 | 13.680+-0.014 | ..... | ..... | ..... | ..... |
| 1891.76 | 30.00 | 13.730+-0.014 | ..... | ..... | ..... | ..... |
| 1933.86 | 30.00 | 13.740+-0.014 | ..... | ..... | ..... | ..... |
| 1975.86 | 30.00 | 13.756+-0.014 | ..... | ..... | ..... | ..... |
| 2018.26 | 30.00 | 13.790+-0.014 | ..... | ..... | ..... | ..... |
| 2060.36 | 30.00 | 13.813+-0.014 | ..... | ..... | ..... | ..... |
| 2102.36 | 30.00 | 13.833+-0.014 | ..... | ..... | ..... | ..... |
| 2144.66 | 30.00 | 13.851+-0.014 | ..... | ..... | ..... | ..... |
| 2187.36 | 30.00 | 13.895+-0.015 | ..... | ..... | ..... | ..... |
| 2229.36 | 30.00 | 13.904+-0.015 | ..... | ..... | ..... | ..... |
| 2271.86 | 30.00 | 13.934+-0.015 | ..... | ..... | ..... | ..... |
| 2314.26 | 30.00 | 13.954+-0.015 | ..... | ..... | ..... | ..... |
| 2356.66 | 30.00 | 13.966+-0.015 | ..... | ..... | ..... | ..... |
| 2398.66 | 30.00 | 13.984+-0.015 | ..... | ..... | ..... | ..... |
| 2440.76 | 30.00 | 14.004+-0.015 | ..... | ..... | ..... | ..... |
| 2483.16 | 30.00 | 14.042+-0.015 | ..... | ..... | ..... | ..... |
| 2525.56 | 30.00 | 14.052+-0.015 | ..... | ..... | ..... | ..... |
| 2568.26 | 30.00 | 14.071+-0.015 | ..... | ..... | ..... | ..... |
| 2611.07 | 30.00 | 14.092+-0.015 | ..... | ..... | ..... | ..... |
| 2653.36 | 30.00 | 14.097+-0.015 | ..... | ..... | ..... | ..... |
| 2696.16 | 30.00 | 14.127+-0.015 | ..... | ..... | ..... | ..... |
| 2738.16 | 30.00 | 14.152+-0.016 | ..... | ..... | ..... | ..... |
| 2780.36 | 30.00 | 14.151+-0.016 | ..... | ..... | ..... | ..... |
| 2822.36 | 30.00 | 14.178+-0.016 | ..... | ..... | ..... | ..... |
| 2864.76 | 30.00 | 14.188+-0.016 | ..... | ..... | ..... | ..... |
| 2907.16 | 30.00 | 14.185+-0.016 | ..... | ..... | ..... | ..... |
| 2949.46 | 30.00 | 14.206+-0.016 | ..... | ..... | ..... | ..... |
| 2991.46 | 30.00 | 14.221+-0.016 | ..... | ..... | ..... | ..... |
| 3033.86 | 30.00 | 14.241+-0.016 | ..... | ..... | ..... | ..... |
| 3075.86 | 30.00 | 14.276+-0.016 | ..... | ..... | ..... | ..... |
| 3118.26 | 30.00 | 14.287+-0.016 | ..... | ..... | ..... | ..... |
| 3160.56 | 30.00 | 14.296+-0.017 | ..... | ..... | ..... | ..... |
| 3203.36 | 30.00 | 14.332+-0.017 | ..... | ..... | ..... | ..... |
| 3245.46 | 30.00 | 14.322+-0.017 | ..... | ..... | ..... | ..... |
| 3288.26 | 30.00 | 14.324+-0.017 | ..... | ..... | ..... | ..... |
| 3330.46 | 30.00 | 14.342+-0.017 | ..... | ..... | ..... | ..... |
| 3372.76 | 30.00 | 14.336+-0.017 | ..... | ..... | ..... | ..... |
| 3415.56 | 30.00 | 14.391+-0.017 | ..... | ..... | ..... | ..... |
| 3457.56 | 30.00 | 14.389+-0.017 | ..... | ..... | ..... | ..... |
| 3499.97 | 30.00 | 14.383+-0.017 | ..... | ..... | ..... | ..... |
| 3542.66 | 30.00 | 14.380+-0.018 | ..... | ..... | ..... | ..... |
| 3585.06 | 30.00 | 14.408+-0.018 | ..... | ..... | ..... | ..... |
| 3627.66 | 30.00 | 14.431+-0.018 | ..... | ..... | ..... | ..... |
| 3669.66 | 30.00 | 14.432+-0.018 | ..... | ..... | ..... | ..... |
| 3711.96 | 30.00 | 14.450+-0.018 | ..... | ..... | ..... | ..... |
| 3754.26 | 30.00 | 14.455+-0.018 | ..... | ..... | ..... | ..... |
| 3796.96 | 30.00 | 14.467+-0.018 | ..... | ..... | ..... | ..... |
| 3839.46 | 30.00 | 14.489+-0.018 | ..... | ..... | ..... | ..... |
| 3881.76 | 30.00 | 14.510+-0.018 | ..... | ..... | ..... | ..... |
| 3924.46 | 30.00 | 14.504+-0.018 | ..... | ..... | ..... | ..... |
| 3966.87 | 30.00 | 14.526+-0.019 | ..... | ..... | ..... | ..... |
| 4008.86 | 30.00 | 14.514+-0.018 | ..... | ..... | ..... | ..... |
| 4051.56 | 30.00 | 14.525+-0.018 | ..... | ..... | ..... | ..... |
| 4093.96 | 30.00 | 14.578+-0.019 | ..... | ..... | ..... | ..... |
| 4136.36 | 30.00 | 14.558+-0.019 | ..... | ..... | ..... | ..... |
| 4178.56 | 30.00 | 14.587+-0.019 | ..... | ..... | ..... | ..... |
| 4221.26 | 30.00 | 14.587+-0.019 | ..... | ..... | ..... | ..... |
| 4263.46 | 30.00 | 14.583+-0.019 | ..... | ..... | ..... | ..... |
| 4305.86 | 30.00 | 14.592+-0.019 | ..... | ..... | ..... | ..... |
| 4348.66 | 30.00 | 14.574+-0.019 | ..... | ..... | ..... | ..... |
| 4391.06 | 30.00 | 14.596+-0.019 | ..... | ..... | ..... | ..... |
| 4433.37 | 30.00 | 14.604+-0.019 | ..... | ..... | ..... | ..... |
| 4476.06 | 30.00 | 14.606+-0.019 | ..... | ..... | ..... | ..... |
| 4518.16 | 30.00 | 14.605+-0.020 | ..... | ..... | ..... | ..... |
| 4560.46 | 30.00 | 14.651+-0.020 | ..... | ..... | ..... | ..... |
| 4602.66 | 30.00 | 14.634+-0.020 | ..... | ..... | ..... | ..... |
| 4645.06 | 30.00 | 14.664+-0.020 | ..... | ..... | ..... | ..... |
| 4687.56 | 30.00 | 14.649+-0.020 | ..... | ..... | ..... | ..... |
| 4730.26 | 30.00 | 14.674+-0.020 | ..... | ..... | ..... | ..... |
| 4772.46 | 30.00 | 14.655+-0.020 | ..... | ..... | ..... | ..... |
| 4814.86 | 30.00 | 14.679+-0.020 | ..... | ..... | ..... | ..... |
| 4857.26 | 30.00 | 14.670+-0.021 | ..... | ..... | ..... | ..... |
| 4899.36 | 30.00 | 14.717+-0.021 | ..... | ..... | ..... | ..... |
| 4941.66 | 30.00 | 14.723+-0.020 | ..... | ..... | ..... | ..... |
| 4983.66 | 30.00 | 14.719+-0.021 | ..... | ..... | ..... | ..... |
| 5026.06 | 30.00 | 14.772+-0.022 | ..... | ..... | ..... | ..... |

---

RAPTOR-T

---

| | | | | | | |
|---|---|---|---|---|---|---|
| 138.20 | 5.00 | ..... | 11.098+-0.019 | 10.822+-0.009 | 10.665+-0.013 | 10.498+-0.025 |
| 149.62 | 5.00 | ..... | 11.179+-0.019 | 10.894+-0.009 | 10.716+-0.013 | 10.565+-0.025 |
| 161.03 | 5.00 | ..... | 11.231+-0.019 | 10.934+-0.010 | 10.800+-0.013 | 10.675+-0.025 |
| 172.44 | 5.00 | ..... | 11.278+-0.019 | 10.974+-0.010 | 10.859+-0.013 | 10.716+-0.025 |
| 183.85 | 5.00 | ..... | 11.263+-0.019 | 11.008+-0.010 | 10.899+-0.013 | 10.780+-0.025 |
| 195.66 | 5.00 | ..... | 11.337+-0.020 | 11.080+-0.010 | 10.955+-0.013 | 10.786+-0.025 |
| 207.58 | 5.00 | ..... | 11.337+-0.020 | 11.108+-0.010 | 10.971+-0.013 | 10.875+-0.025 |
| 218.99 | 5.00 | ..... | 11.447+-0.020 | 11.159+-0.010 | 11.037+-0.013 | 10.902+-0.025 |
| 230.81 | 5.00 | ..... | 11.451+-0.020 | 11.173+-0.010 | 11.088+-0.013 | 10.893+-0.026 |
| 247.24 | 10.00 | ..... | 11.582+-0.019 | 11.289+-0.009 | 11.178+-0.013 | 10.986+-0.025 |
| 260.61 | 10.00 | ..... | 11.585+-0.019 | 11.319+-0.009 | 11.178+-0.013 | 11.027+-0.024 |
| 273.60 | 10.00 | ..... | 11.631+-0.019 | 11.359+-0.009 | 11.237+-0.013 | 11.127+-0.024 |
| 286.97 | 10.00 | ..... | 11.687+-0.019 | 11.406+-0.009 | 11.290+-0.013 | 11.139+-0.024 |
| 299.96 | 10.00 | ..... | 11.745+-0.019 | 11.467+-0.009 | 11.337+-0.013 | 11.200+-0.025 |
| 312.98 | 10.00 | ..... | 11.822+-0.019 | 11.520+-0.009 | 11.414+-0.013 | 11.238+-0.025 |
| 325.94 | 10.00 | ..... | 11.867+-0.019 | 11.555+-0.010 | 11.456+-0.013 | 11.300+-0.025 |
| 338.93 | 10.00 | ..... | 11.909+-0.019 | 11.606+-0.010 | 11.497+-0.013 | 11.366+-0.025 |
| 352.30 | 10.00 | ..... | 11.933+-0.019 | 11.672+-0.010 | 11.545+-0.013 | 11.389+-0.025 |
| 365.69 | 10.00 | ..... | 12.032+-0.020 | 11.709+-0.010 | 11.604+-0.013 | 11.436+-0.025 |
| 379.07 | 10.00 | ..... | 12.064+-0.020 | 11.770+-0.010 | 11.620+-0.013 | 11.469+-0.025 |
| 392.46 | 10.00 | ..... | 12.088+-0.020 | 11.801+-0.010 | 11.667+-0.013 | 11.527+-0.025 |
| 405.83 | 10.00 | ..... | 12.135+-0.020 | 11.842+-0.010 | 11.716+-0.013 | 11.571+-0.025 |
| 419.01 | 10.00 | ..... | 12.171+-0.020 | 11.878+-0.010 | 11.735+-0.013 | 11.636+-0.026 |
| 431.99 | 10.00 | ..... | 12.212+-0.020 | 11.925+-0.010 | 11.803+-0.013 | 11.606+-0.026 |
| 445.17 | 10.00 | ..... | 12.261+-0.020 | 11.957+-0.010 | 11.825+-0.013 | 11.671+-0.026 |
| 458.55 | 10.00 | ..... | 12.283+-0.020 | 11.974+-0.010 | 11.870+-0.014 | 11.729+-0.026 |
| 472.13 | 10.00 | ..... | 12.302+-0.020 | 12.028+-0.010 | 11.883+-0.014 | 11.718+-0.026 |
| 485.10 | 10.00 | ..... | 12.342+-0.020 | 12.058+-0.010 | 11.925+-0.014 | 11.757+-0.026 |
| 498.08 | 10.00 | ..... | 12.399+-0.020 | 12.101+-0.010 | 11.960+-0.014 | 11.779+-0.026 |
| 525.15 | 30.00 | ..... | 12.458+-0.019 | 12.153+-0.009 | 12.016+-0.013 | 11.864+-0.024 |
| 567.05 | 30.00 | ..... | 12.526+-0.019 | 12.257+-0.009 | 12.119+-0.013 | ..... |
| 608.32 | 30.00 | ..... | 12.610+-0.019 | 12.324+-0.009 | 12.188+-0.013 | ..... |
| 650.06 | 30.00 | ..... | 12.696+-0.019 | 12.407+-0.009 | 12.270+-0.013 | ..... |
| 691.73 | 30.00 | ..... | 12.763+-0.019 | 12.464+-0.009 | 12.335+-0.013 | ..... |
| 733.47 | 30.00 | ..... | 12.822+-0.019 | 12.528+-0.009 | 12.396+-0.013 | ..... |
| 774.78 | 30.00 | ..... | 12.888+-0.020 | 12.589+-0.010 | 12.481+-0.013 | ..... |
| 816.03 | 30.00 | ..... | 12.949+-0.020 | 12.670+-0.010 | 12.529+-0.013 | ..... |
| 857.68 | 30.00 | ..... | 13.021+-0.020 | 12.749+-0.010 | 12.579+-0.013 | ..... |
| 898.98 | 30.00 | ..... | 13.073+-0.020 | 12.800+-0.010 | 12.659+-0.013 | ..... |
| 940.65 | 30.00 | ..... | 13.153+-0.020 | 12.877+-0.010 | 12.713+-0.013 | ..... |
| 982.35 | 30.00 | ..... | 13.241+-0.020 | 12.941+-0.010 | 12.780+-0.014 | ..... |
| 1023.70 | 30.00 | ..... | 13.280+-0.020 | 12.996+-0.010 | 12.873+-0.014 | ..... |
| 1065.60 | 30.00 | ..... | 13.365+-0.020 | 13.063+-0.010 | 12.901+-0.014 | ..... |
| 1107.42 | 30.00 | ..... | 13.391+-0.020 | 13.105+-0.010 | 12.978+-0.014 | ..... |
| 1148.82 | 30.00 | ..... | 13.453+-0.020 | 13.141+-0.010 | 13.029+-0.014 | ..... |
| 1190.49 | 30.00 | ..... | 13.486+-0.020 | 13.199+-0.010 | 13.052+-0.014 | ..... |
| 1231.82 | 30.00 | ..... | 13.539+-0.020 | 13.234+-0.011 | 13.117+-0.014 | ..... |
| 1273.43 | 30.00 | ..... | 13.588+-0.021 | 13.289+-0.011 | 13.162+-0.015 | ..... |
| 1315.24 | 30.00 | ..... | 13.632+-0.021 | 13.330+-0.011 | 13.213+-0.015 | ..... |
| 1357.05 | 30.00 | ..... | 13.651+-0.021 | 13.371+-0.011 | 13.253+-0.015 | ..... |
| 1399.05 | 30.00 | ..... | 13.683+-0.021 | 13.418+-0.011 | 13.271+-0.016 | ..... |
| 1440.96 | 30.00 | ..... | 13.739+-0.021 | 13.462+-0.011 | 13.324+-0.016 | ..... |
| 1483.47 | 30.00 | ..... | 13.766+-0.021 | 13.493+-0.011 | 13.351+-0.016 | ..... |
| 1525.18 | 30.00 | ..... | 13.787+-0.021 | 13.518+-0.011 | 13.377+-0.016 | ..... |
| 1566.88 | 30.00 | ..... | 13.830+-0.021 | 13.543+-0.011 | 13.428+-0.016 | ..... |
| 1608.69 | 30.00 | ..... | 13.856+-0.021 | 13.593+-0.011 | 13.438+-0.016 | ..... |
| 1649.99 | 30.00 | ..... | 13.886+-0.021 | 13.596+-0.011 | 13.470+-0.016 | ..... |
| 1691.60 | 30.00 | ..... | 13.934+-0.021 | 13.639+-0.012 | 13.514+-0.017 | ..... |
| 1733.71 | 30.00 | ..... | 13.961+-0.021 | 13.675+-0.012 | 13.534+-0.017 | ..... |
| 1775.52 | 30.00 | ..... | 13.984+-0.021 | 13.675+-0.012 | 13.530+-0.017 | ..... |
| 1816.78 | 30.00 | ..... | 13.989+-0.022 | 13.711+-0.012 | 13.580+-0.018 | ..... |
| 1858.48 | 30.00 | ..... | 14.007+-0.022 | 13.731+-0.012 | 13.603+-0.018 | ..... |
| 1900.33 | 30.00 | ..... | 14.039+-0.022 | 13.756+-0.012 | 13.649+-0.018 | ..... |
| 1941.84 | 30.00 | ..... | 14.064+-0.022 | 13.781+-0.013 | 13.655+-0.018 | ..... |
| 1983.54 | 30.00 | ..... | 14.088+-0.022 | 13.825+-0.013 | 13.704+-0.018 | ..... |
| 2024.94 | 30.00 | ..... | 14.108+-0.022 | 13.845+-0.013 | 13.716+-0.018 | ..... |
| 2066.60 | 30.00 | ..... | 14.145+-0.022 | 13.863+-0.013 | 13.737+-0.018 | ..... |
| 2108.86 | 30.00 | ..... | 14.149+-0.022 | 13.872+-0.013 | 13.774+-0.018 | ..... |
| 2150.16 | 30.00 | ..... | 14.180+-0.022 | 13.914+-0.013 | 13.754+-0.018 | ..... |
| 2191.82 | 30.00 | ..... | 14.185+-0.022 | 13.921+-0.013 | 13.776+-0.018 | ..... |
| 2233.87 | 30.00 | ..... | 14.213+-0.023 | 13.947+-0.013 | 13.809+-0.019 | ..... |
| 2275.28 | 30.00 | ..... | 14.243+-0.023 | 13.972+-0.013 | 13.867+-0.020 | ..... |
| 2316.68 | 30.00 | ..... | 14.262+-0.023 | 14.005+-0.013 | 13.872+-0.020 | ..... |
| 2358.59 | 30.00 | ..... | 14.270+-0.023 | 13.998+-0.013 | 13.885+-0.020 | ..... |
| 2400.39 | 30.00 | ..... | 14.317+-0.024 | 14.039+-0.013 | 13.901+-0.020 | ..... |
| 2442.10 | 30.00 | ..... | 14.337+-0.024 | 14.048+-0.013 | 13.894+-0.020 | ..... |
| 2483.80 | 30.00 | ..... | 14.327+-0.024 | 14.051+-0.014 | 13.941+-0.021 | ..... |
| 2525.61 | 30.00 | ..... | 14.369+-0.024 | 14.091+-0.014 | 13.934+-0.021 | ..... |
| 2567.27 | 30.00 | ..... | 14.388+-0.024 | 14.101+-0.014 | 13.986+-0.021 | ..... |
| 2608.98 | 30.00 | ..... | 14.366+-0.024 | 14.145+-0.014 | 14.026+-0.022 | ..... |
| 2650.73 | 30.00 | ..... | 14.422+-0.024 | 14.119+-0.014 | 14.029+-0.022 | ..... |
| 2692.53 | 30.00 | ..... | 14.428+-0.025 | 14.173+-0.015 | 14.048+-0.022 | ..... |
| 2734.20 | 30.00 | ..... | 14.457+-0.025 | 14.175+-0.015 | 14.030+-0.022 | ..... |
| 2775.54 | 30.00 | ..... | 14.445+-0.025 | 14.208+-0.015 | 14.072+-0.022 | ..... |
| 2817.20 | 30.00 | ..... | 14.494+-0.025 | 14.197+-0.015 | 14.094+-0.022 | ..... |
| 2858.51 | 30.00 | ..... | 14.498+-0.025 | 14.216+-0.015 | 14.095+-0.022 | ..... |
| 2900.66 | 30.00 | ..... | 14.524+-0.025 | 14.258+-0.016 | 14.061+-0.022 | ..... |
| 2942.67 | 30.00 | ..... | 14.530+-0.025 | 14.268+-0.016 | 14.111+-0.022 | ..... |
| 2984.57 | 30.00 | ..... | 14.564+-0.026 | 14.253+-0.016 | 14.149+-0.023 | ..... |
| 3025.88 | 30.00 | ..... | 14.567+-0.026 | 14.268+-0.016 | 14.136+-0.023 | ..... |
| 3067.48 | 30.00 | ..... | 14.552+-0.026 | 14.289+-0.016 | 14.175+-0.024 | ..... |
| 3109.39 | 30.00 | ..... | 14.590+-0.026 | 14.289+-0.016 | 14.188+-0.024 | ..... |
| 3150.99 | 30.00 | ..... | 14.586+-0.026 | 14.320+-0.017 | 14.319+-0.025 | ..... |
| 3192.50 | 30.00 | ..... | 14.586+-0.027 | 14.366+-0.017 | 14.194+-0.024 | ..... |
| 3234.51 | 30.00 | ..... | 14.606+-0.026 | 14.371+-0.017 | 14.248+-0.025 | ..... |
| 3276.52 | 30.00 | ..... | 14.605+-0.027 | 14.347+-0.017 | 14.242+-0.025 | ..... |
| 3317.92 | 30.00 | ..... | 14.660+-0.027 | 14.377+-0.017 | 14.228+-0.025 | ..... |
| 3359.58 | 30.00 | ..... | 14.660+-0.027 | 14.394+-0.017 | 14.312+-0.026 | ..... |
| 3401.29 | 30.00 | ..... | 14.673+-0.028 | 14.397+-0.017 | 14.274+-0.025 | ..... |
| 3443.00 | 30.00 | ..... | 14.651+-0.027 | 14.443+-0.017 | 14.431+-0.027 | ..... |
| 3484.70 | 30.00 | ..... | 14.701+-0.028 | 14.438+-0.017 | 14.309+-0.025 | ..... |
| 3526.55 | 30.00 | ..... | 14.705+-0.028 | 14.417+-0.017 | 14.311+-0.026 | ..... |
| 3568.56 | 30.00 | ..... | 14.748+-0.028 | 14.458+-0.017 | 14.455+-0.028 | ..... |
| 3610.37 | 30.00 | ..... | 14.704+-0.028 | 14.466+-0.017 | 14.361+-0.027 | ..... |
| 3651.66 | 30.00 | ..... | 14.740+-0.028 | 14.490+-0.018 | ..... | ..... |
| 3693.17 | 30.00 | ..... | 14.826+-0.030 | 14.509+-0.018 | ..... | ..... |
| 3735.08 | 30.00 | ..... | 14.773+-0.028 | 14.513+-0.018 | ..... | ..... |
| 3776.79 | 30.00 | ..... | 14.753+-0.028 | 14.498+-0.018 | ..... | ..... |
| 3818.70 | 30.00 | ..... | 14.751+-0.028 | 14.503+-0.018 | ..... | ..... |
| 3860.10 | 30.00 | ..... | 14.759+-0.029 | 14.523+-0.018 | ..... | ..... |
| 3901.50 | 30.00 | ..... | 14.789+-0.029 | 14.543+-0.019 | ..... | ..... |
| 3943.61 | 30.00 | ..... | 14.759+-0.028 | 14.531+-0.018 | ..... | ..... |
| 3985.32 | 30.00 | ..... | 14.819+-0.030 | 14.562+-0.019 | ..... | ..... |
| 4027.12 | 30.00 | ..... | 14.823+-0.030 | 14.546+-0.019 | ..... | ..... |
| 4068.78 | 30.00 | ..... | 14.843+-0.031 | 14.568+-0.019 | ..... | ..... |
| 4110.33 | 30.00 | ..... | 14.880+-0.031 | 14.595+-0.020 | ..... | ..... |
| 4152.34 | 30.00 | ..... | 14.843+-0.031 | 14.576+-0.019 | ..... | ..... |
| 4194.05 | 30.00 | ..... | 14.842+-0.031 | 14.590+-0.020 | ..... | ..... |
| 4237.97 | 30.00 | ..... | 14.885+-0.031 | 14.615+-0.020 | ..... | ..... |
| 4279.78 | 30.00 | ..... | 14.873+-0.031 | 14.643+-0.020 | ..... | ..... |
| 4321.08 | 30.00 | ..... | 14.868+-0.031 | 14.651+-0.020 | ..... | ..... |
| 4362.49 | 30.00 | ..... | 14.913+-0.031 | 14.619+-0.020 | ..... | ..... |
| 4404.39 | 30.00 | ..... | 14.950+-0.032 | 14.635+-0.020 | ..... | ..... |
| 4446.10 | 30.00 | ..... | 14.893+-0.031 | 14.656+-0.021 | ..... | ..... |
| 4487.47 | 30.00 | ..... | 14.950+-0.032 | 14.694+-0.021 | ..... | ..... |
| 4529.11 | 30.00 | ..... | 14.913+-0.031 | 14.705+-0.021 | ..... | ..... |
| 4571.01 | 30.00 | ..... | 14.950+-0.033 | 14.691+-0.021 | ..... | ..... |
| 4613.02 | 30.00 | ..... | 14.971+-0.033 | 14.691+-0.021 | ..... | ..... |
| 4654.32 | 30.00 | ..... | 14.990+-0.034 | 14.694+-0.022 | ..... | ..... |
| 4695.73 | 30.00 | ..... | 14.996+-0.034 | 14.732+-0.022 | ..... | ..... |
| 4737.33 | 30.00 | ..... | 14.978+-0.033 | 14.710+-0.022 | ..... | ..... |
| 4778.94 | 30.00 | ..... | 14.954+-0.033 | 14.707+-0.022 | ..... | ..... |
| 4820.34 | 30.00 | ..... | 14.965+-0.033 | 14.754+-0.023 | ..... | ..... |
| 4862.15 | 30.00 | ..... | 14.978+-0.033 | 14.722+-0.022 | ..... | ..... |
| 4903.65 | 30.00 | ..... | 14.968+-0.034 | 14.743+-0.023 | ..... | ..... |
| 4944.95 | 30.00 | ..... | 15.035+-0.035 | 14.752+-0.023 | ..... | ..... |
| 4986.76 | 30.00 | ..... | 15.027+-0.035 | 14.795+-0.023 | ..... | ..... |
| 5028.77 | 30.00 | ..... | 15.013+-0.035 | 14.785+-0.024 | ..... | ..... |
| 5070.48 | 30.00 | ..... | 15.026+-0.035 | 14.808+-0.024 | ..... | ..... |
| 5112.19 | 30.00 | ..... | 15.019+-0.035 | 14.810+-0.024 | ..... | ..... |
| 5153.79 | 30.00 | ..... | 15.064+-0.036 | 14.824+-0.025 | ..... | ..... |
| 5195.70 | 30.00 | ..... | 15.072+-0.036 | 14.802+-0.024 | ..... | ..... |
| 5236.99 | 30.00 | ..... | 15.071+-0.036 | 14.818+-0.024 | ..... | ..... |
| 5278.26 | 30.00 | ..... | 15.063+-0.036 | 14.867+-0.025 | ..... | ..... |
| 5320.31 | 30.00 | ..... | 15.051+-0.036 | 14.833+-0.024 | ..... | ..... |
| 5362.02 | 30.00 | ..... | 15.090+-0.037 | 14.818+-0.025 | ..... | ..... |
| 5403.82 | 30.00 | ..... | 15.040+-0.036 | 14.877+-0.026 | ..... | ..... |
| 5445.63 | 30.00 | ..... | 15.123+-0.038 | 14.886+-0.026 | ..... | ..... |
| 5487.53 | 30.00 | ..... | 15.110+-0.037 | 14.889+-0.026 | ..... | ..... |
| 5528.94 | 30.00 | ..... | 15.165+-0.039 | 14.868+-0.026 | ..... | ..... |
| 5570.24 | 30.00 | ..... | 15.099+-0.038 | 14.866+-0.026 | ..... | ..... |
| 5611.90 | 30.00 | ..... | 15.074+-0.036 | 14.873+-0.026 | ..... | ..... |
| 5653.65 | 30.00 | ..... | 15.159+-0.039 | 14.888+-0.026 | ..... | ..... |
| 5695.76 | 30.00 | ..... | 15.186+-0.039 | 14.901+-0.026 | ..... | ..... |
| 5737.22 | 30.00 | ..... | 15.189+-0.040 | 14.905+-0.027 | ..... | ..... |
| 5779.17 | 30.00 | ..... | 15.185+-0.040 | 14.927+-0.027 | ..... | ..... |
| 5820.98 | 30.00 | ..... | 15.192+-0.040 | 14.927+-0.028 | ..... | ..... |
| 5862.68 | 30.00 | ..... | 15.163+-0.040 | 14.910+-0.027 | ..... | ..... |
| 5904.39 | 30.00 | ..... | 15.194+-0.041 | 14.891+-0.027 | ..... | ..... |
| 5946.10 | 30.00 | ..... | 15.194+-0.041 | 14.943+-0.028 | ..... | ..... |
| 5987.80 | 30.00 | ..... | 15.183+-0.041 | 14.975+-0.028 | ..... | ..... |
| 6029.11 | 30.00 | ..... | 15.177+-0.041 | 14.909+-0.028 | ..... | ..... |
| 6070.91 | 30.00 | ..... | 15.169+-0.041 | 14.970+-0.028 | ..... | ..... |
| 6112.58 | 30.00 | ..... | 15.125+-0.039 | 14.988+-0.028 | ..... | ..... |
| 6154.28 | 30.00 | ..... | 15.231+-0.042 | 14.986+-0.028 | ..... | ..... |
| 6195.99 | 30.00 | ..... | 15.204+-0.042 | 14.983+-0.028 | ..... | ..... |
| 6237.43 | 30.00 | ..... | 15.212+-0.042 | 14.977+-0.028 | ..... | ..... |
| 6279.34 | 30.00 | ..... | 15.238+-0.044 | 14.965+-0.028 | ..... | ..... |
| 6320.74 | 30.00 | ..... | 15.256+-0.044 | 14.986+-0.029 | ..... | ..... |
| 6362.24 | 30.00 | ..... | 15.257+-0.045 | 15.036+-0.030 | ..... | ..... |
| 6403.95 | 30.00 | ..... | 15.208+-0.042 | 15.007+-0.029 | ..... | ..... |
| 6445.76 | 30.00 | ..... | 15.258+-0.044 | 14.984+-0.029 | ..... | ..... |
| 6487.46 | 30.00 | ..... | 15.215+-0.042 | 15.035+-0.030 | ..... | ..... |
| 6529.27 | 30.00 | ..... | 15.367+-0.048 | 15.002+-0.029 | ..... | ..... |
| 6570.97 | 30.00 | ..... | 15.256+-0.044 | 15.044+-0.031 | ..... | ..... |
| 6612.28 | 30.00 | ..... | 15.329+-0.049 | 15.057+-0.032 | ..... | ..... |
| 6653.58 | 30.00 | ..... | 15.242+-0.045 | 15.080+-0.032 | ..... | ..... |
| 6695.29 | 30.00 | ..... | 15.245+-0.044 | 15.059+-0.031 | ..... | ..... |
| 6736.54 | 30.00 | ..... | 15.322+-0.047 | 15.057+-0.032 | ..... | ..... |
| 6777.89 | 30.00 | ..... | 15.295+-0.047 | 15.037+-0.032 | ..... | ..... |
| 6819.29 | 30.00 | ..... | 15.340+-0.048 | 15.053+-0.032 | ..... | ..... |
| 6861.10 | 30.00 | ..... | 15.276+-0.046 | 15.034+-0.031 | ..... | ..... |
| 6903.00 | 30.00 | ..... | 15.294+-0.047 | 15.080+-0.033 | ..... | ..... |
| 6944.82 | 30.00 | ..... | 15.294+-0.047 | 15.024+-0.031 | ..... | ..... |
| 6986.47 | 30.00 | ..... | 15.224+-0.045 | 15.117+-0.033 | ..... | ..... |
| 7028.33 | 30.00 | ..... | 15.355+-0.049 | 15.148+-0.035 | ..... | ..... |
| 7069.73 | 30.00 | ..... | 15.372+-0.050 | 15.066+-0.033 | ..... | ..... |
| 7111.44 | 30.00 | ..... | 15.348+-0.051 | 15.132+-0.034 | ..... | ..... |
| 7152.94 | 30.00 | ..... | 15.340+-0.051 | 15.100+-0.034 | ..... | ..... |
| 7194.75 | 30.00 | ..... | 15.283+-0.048 | 15.076+-0.033 | ..... | ..... |
| 7236.45 | 30.00 | ..... | 15.378+-0.051 | 15.130+-0.035 | ..... | ..... |
| 7278.16 | 30.00 | ..... | 15.354+-0.051 | 15.124+-0.035 | ..... | ..... |
| 7319.96 | 30.00 | ..... | 15.425+-0.053 | 15.161+-0.036 | ..... | ..... |
| 7361.67 | 30.00 | ..... | 15.396+-0.053 | 15.182+-0.037 | ..... | ..... |
| 7403.37 | 30.00 | ..... | 15.333+-0.051 | 15.151+-0.036 | ..... | ..... |
| 7445.39 | 30.00 | ..... | 15.458+-0.055 | 15.185+-0.037 | ..... | ..... |
| 7487.39 | 30.00 | ..... | 15.428+-0.055 | 15.106+-0.035 | ..... | ..... |
| 7529.30 | 30.00 | ..... | 15.337+-0.052 | 15.166+-0.036 | ..... | ..... |
| 7570.90 | 30.00 | ..... | 15.475+-0.058 | 15.173+-0.037 | ..... | ..... |

**Table S1.** RAPTOR observations of GRB130427A. $T_{mid}$ is the mid-exposure time in seconds since the GBM trigger (07:47:06.42 UT). The RAPTOR-T data were calibrated using field stars from the SDSS DR9 catalog (6). The RAPTOR All-Sky Monitor data were calibrated against the Tycho-2 Catalog SDSS r' estimates (5).

**References:**
1. P. Meszaros and M. Rees, *Astrophysical J.*, **476**, 232 (1997).
2. R. Blandford and C. McKee, *Physics of Fluids*, vol 19, no 8, 1130 (1976).
3. E. Bertin, and S. Arnouts, *Astron. Astrophys. Sup.* **117**, 393 (1996).
4. E. Hog, et al., *Astron. Astrophys.* **355**, L27 (2000).
5. E. Pickles and E. Depagne, *Pub. Astron. Soc. Pac.,* **122**, issue 898, 1437 (2010).
6. C. Ahn et al., *Astrophys. J. Supp.* **203**, 21 (2012).
7. Y. C. Pei, *Astrophys. J.*, **395**, 130 (1992).
8. D. J. Schlegel et al., Astrophys. J., **500**, 525 (1998).
9. T. Laskar et al., ArXiv:1305.2453 (2013).